\documentclass{PoS}
\newcommand{\nuc}[2]{\ensuremath{\mathrm{^{#1}#2}}}

\usepackage{verbatim}
\title{Nucleosynthetic post-processing of Type Ia supernovae with variable tracer masses}

\ShortTitle{Nucleosynthetic post-processing of Type Ia supernovae with
  variable tracer masses}

\author{\speaker{Ivo Rolf Seitenzahl}%
         \thanks{Corresponding author.}\\
        Max-Planck-Institut f{\"u}r Astrophysik, Karl-Schwarzschild-Str.~1, D-85748 Garching, Germany\\
        E-mail: \email{irs@mpa-garching.mpg.de}}

\author{Friedrich R\"opke\\
        Max-Planck-Institut f{\"u}r Astrophysik, Karl-Schwarzschild-Str.~1, D-85748 Garching, Germany\\
        E-mail: \email{fritz@mpa-garching.mpg.de}}

\author{R\"udiger Pakmor\\
        Max-Planck-Institut f{\"u}r Astrophysik, Karl-Schwarzschild-Str.~1, D-85748 Garching, Germany\\
        E-mail: \email{rpakmor@mpa-garching.mpg.de}}

\author{Michael Fink\\
        Max-Planck-Institut f{\"u}r Astrophysik, Karl-Schwarzschild-Str.~1, D-85748 Garching, Germany\\
        E-mail: \email{mfink@mpa-garching.mpg.de}}

      \abstract{The post-processing of passively advected Lagrangian
        tracer particles \cite{nagataki1997a} is still the most common
        way for obtaining detailed nucleosynthetic yield predictions
        of Type Ia supernova (SN Ia) hydrodynamical
        simulations. Historically, tracer particles of constant mass
        are employed. However, intermediate mass elements, such as
        e.g. Ne, Mg, Al, or Si, are typically synthesized in the outer
        layers of SNe Ia, where due to the lower initial density a
        constant mass tracer distribution results in poor resolution
        of the spatial morphology of the abundance distribution. We
        show how to alleviate this problem with a suitably chosen
        distribution of variable tracer particle masses. We
        also present results of the convergence of integrated
        nucleosynthetic yields with increasing tracer particle number. We find
        that the yields of the most abundant species (mass fraction >
        $10^{-5}$) are reasonably well predicted for a tracer number
        as small as 32 per axis and direction. Convergence for
        isotopes produced in regions where a constant tracer mass
        implementation results in poor spatial resolution can be
        improved by suitably choosing tracers of variable mass.}

\FullConference{11th Symposium on Nuclei in the Cosmos\\
  19-23 July 2010 \\
  Heidelberg, Germany.}

\begin{document}
\section{Introduction}
Type Ia supernovae (SN Ia) are believed to be thermonuclear explosions
of white dwarf stars. The relative abundances of nuclei synthesized in
SN Ia explosions are dependent on the explosion model and play an
important role in understanding the chemical evolution of our Galaxy
(e.g. \cite{matteucci2009a}). Multi-dimensional numerical simulations
of SN Ia explosions have been carried out by several groups for a
range of different explosion models (see
e.g. \cite{roepke2007c,roepke2007b,bravo2008a,jordan2008a,meakin2009a,pakmor2010a,fink2010a,sim2010a}).
Synthetic light curves and spectra not only depend sensitively on the
amount and location of the radioactive nuclei that decay and reheat
the ejecta (such as e.g. \nuc{56}{Ni} or \nuc{57}{Ni}) but also on the
amount and location of many other elements such as e.g. Mg, Ca, Ti or
Cr, whose (partially ionized) atoms interact with the radiation. For
any explosion model, detailed predictions of the isotopic composition
of the ejecta are therefore needed.  This is commonly done with the
tracer particle method
(e.g. \cite{travaglio2004a,brown2005a,roepke2006a,fink2010a,maeda2010a}),
for which a large number of tracer particles is placed into the
star. The particles are carried along by the flow, recording the local
thermodynamic conditions as a function of time. These ``trajectories''
are then post-processed with a nuclear reaction network and the
resulting yields are assigned to the tracer particles final position
and velocity, weighted by the mass the particle
represents. Historically, tracer particles of equal (constant) mass
are used throughout.

\section{Variable mass tracer particle method}
\begin{figure}[h]
  \centering
  \includegraphics[width=4in]{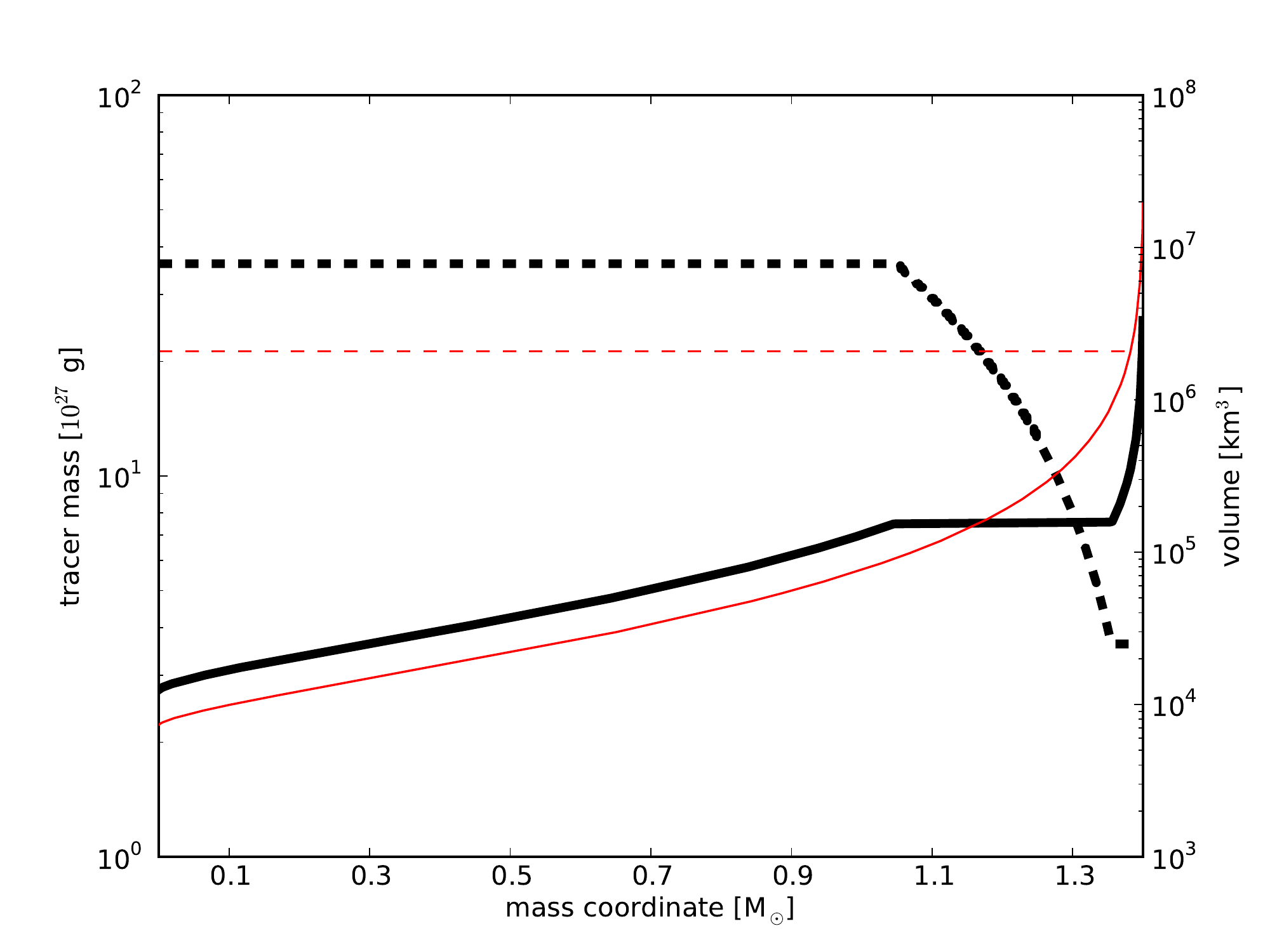}
  \caption{Shown are the mass (dashed lines) and volume (solid lines)
    represented by a tracer particle as a function of its initial
    position in the mass coordinate.  Thin (red) lines are for
    constant tracer particle mass. Thick (black) lines are for
    variable tracer particle mass.}
  \label{fig1}
\end{figure}

\begin{figure}
  \centering
  \begin{tabular}{cc}
    \includegraphics[width=3in]{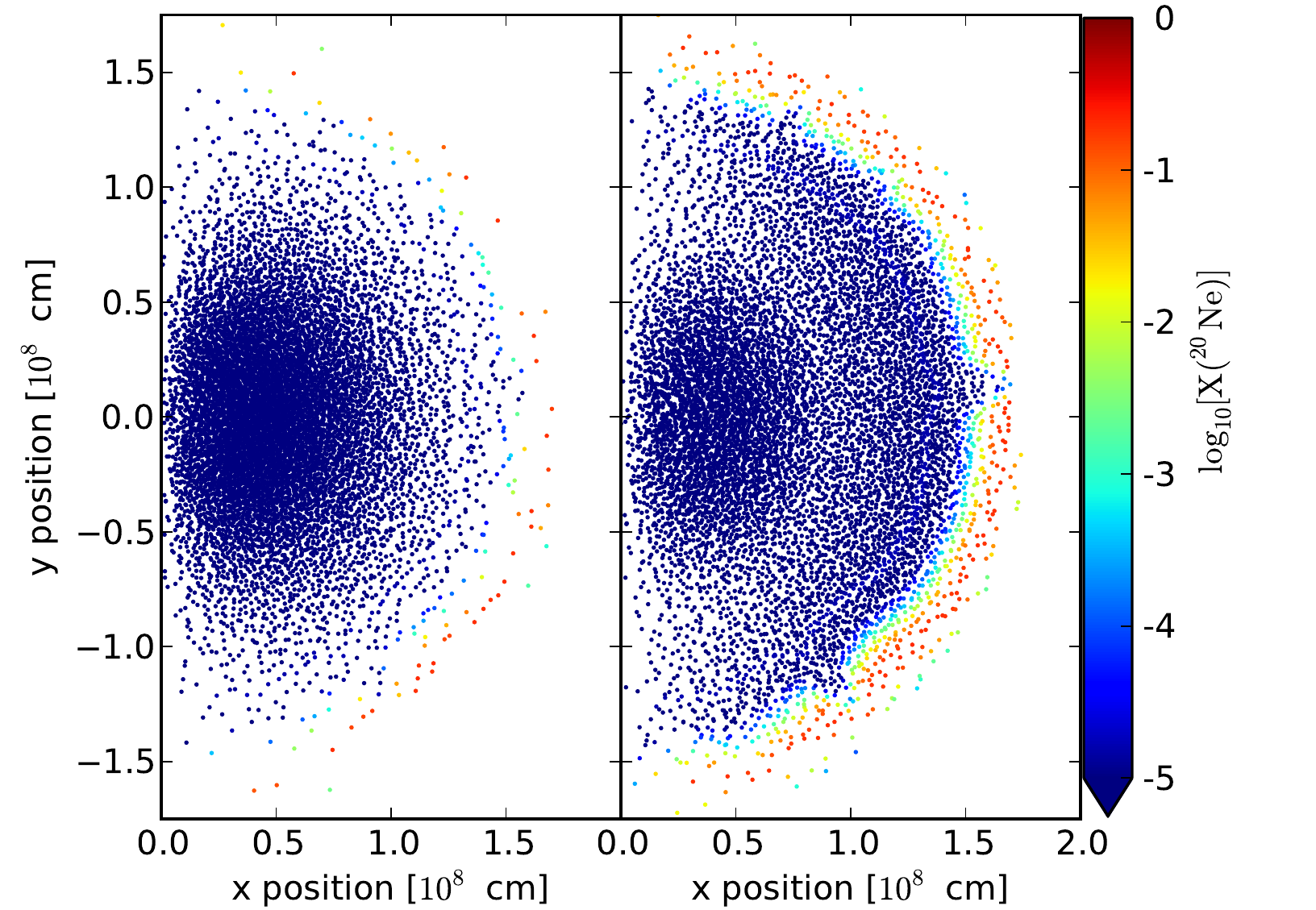} &
    \includegraphics[width=3in]{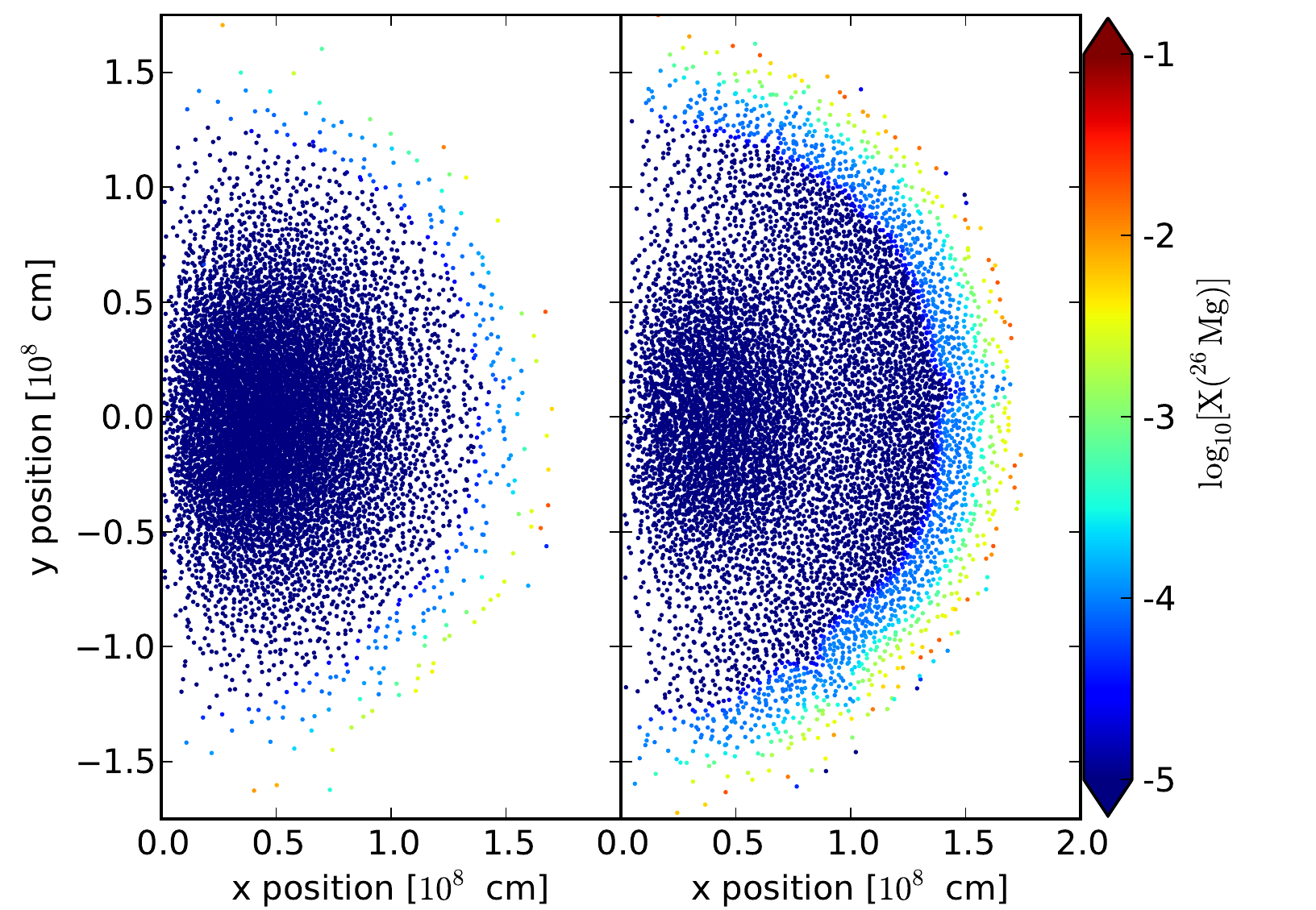} \\
    \includegraphics[width=3in]{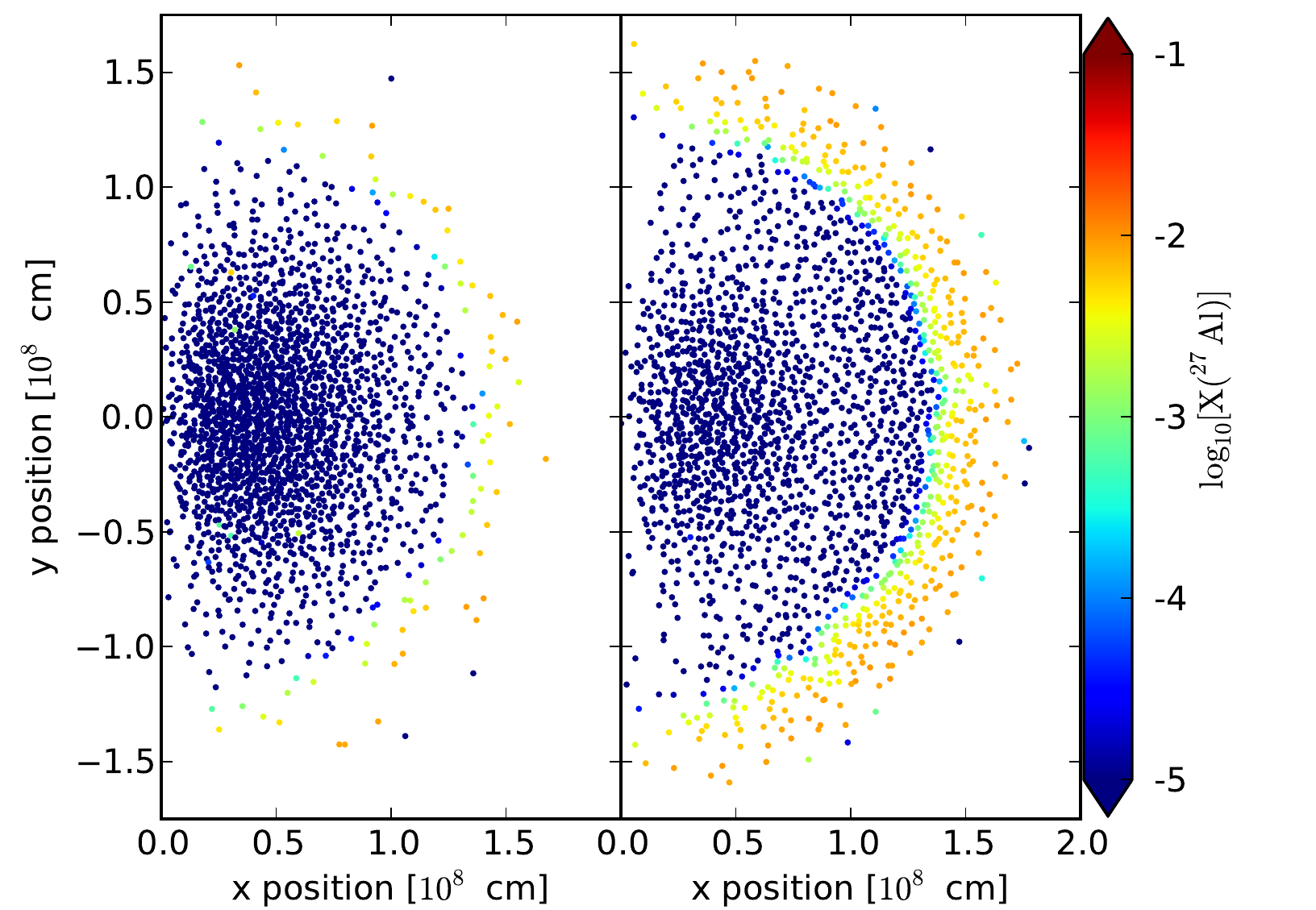} &
    \includegraphics[width=3in]{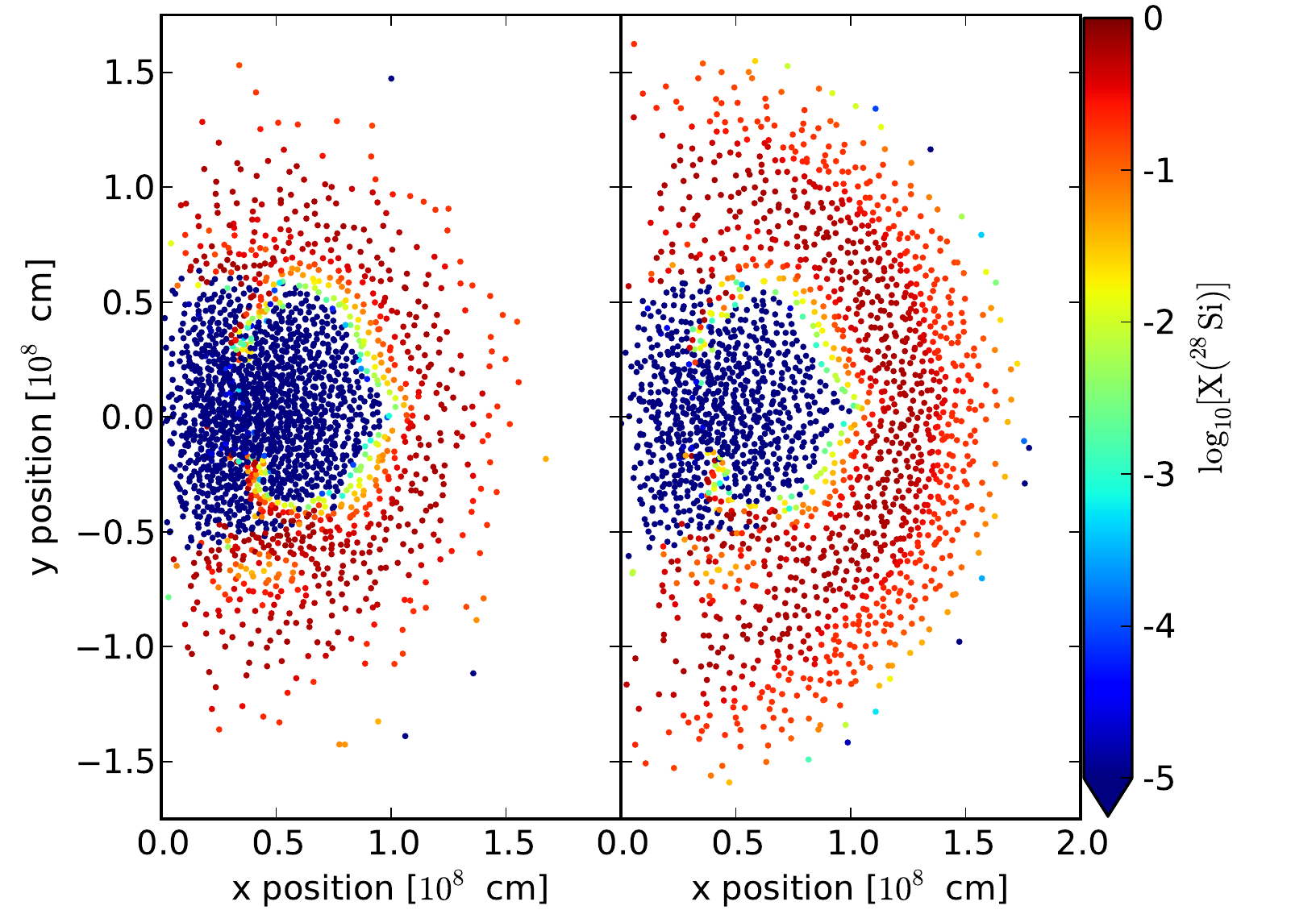}
  \end{tabular}
  \caption{Initial spatial distribution of constant (left side of each
    panel) and variable (right side of each panel) mass tracer
    particles colored by the final mass fraction of \nuc{20}{Ne},
    \nuc{26}{Mg}, \nuc{27}{Al} and \nuc{28}{Si} respectively after
    freeze-out ($t=10~\mathrm{s}$). The variable tracer mass
    distributions better resolve the morphology of these isotopes.}
  \label{fig2}
\end{figure}

\begin{figure*}
  \centering
  \includegraphics[height=7.5in]{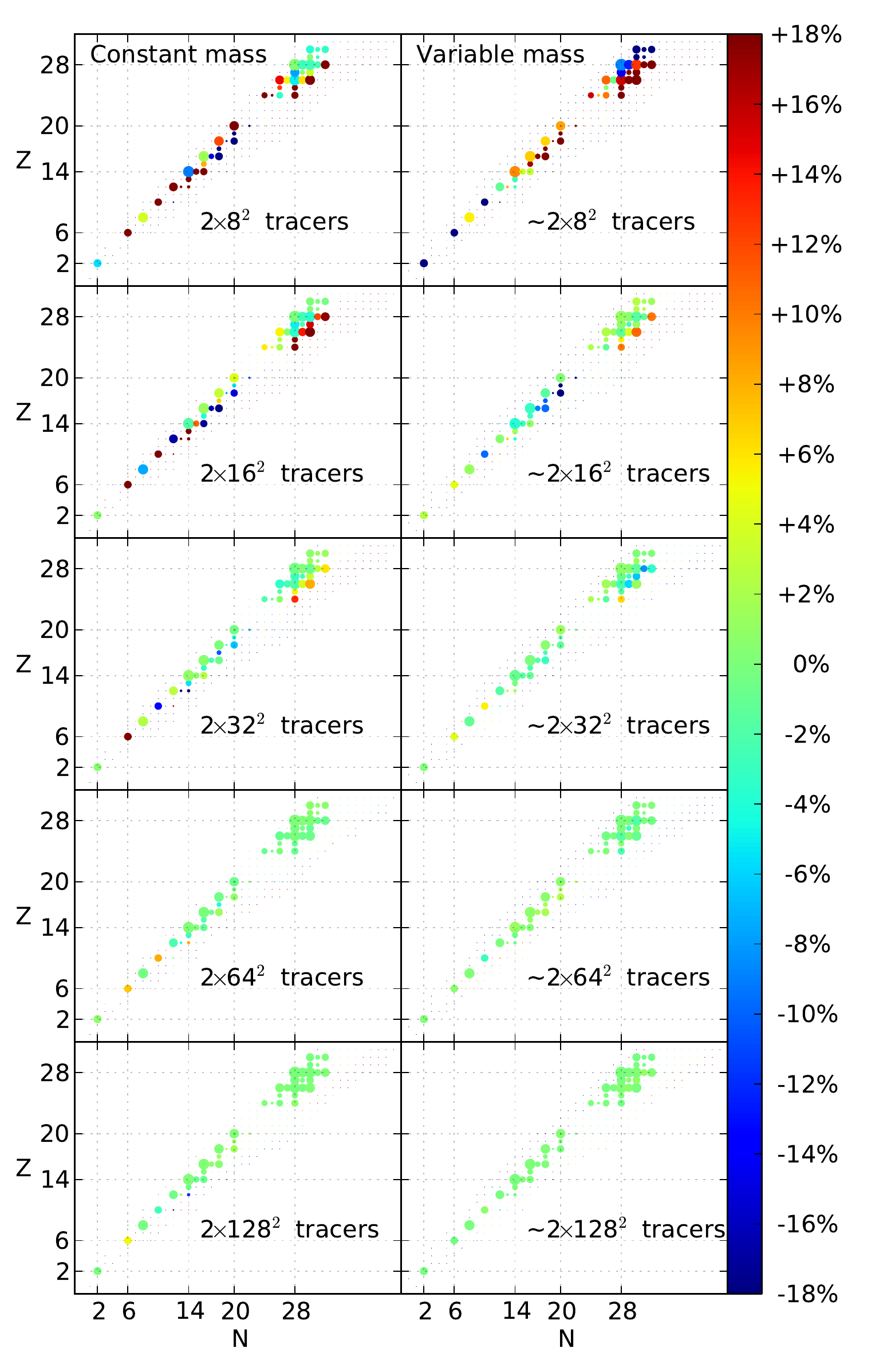}
  \caption{Final ($t=10~\mathrm{s}$) nuclide mass fraction differences
    $\Big[\frac{X_i(2 \times N^2)-X_i(2 \times 256^2)}{X_i(2 \times
      256^2)}\Big]$ in percent for a sequence of increasing total
    tracer particle number compared with the highest resolved case
    containing $2 \times 256^2$ tracer particles.  The underlying
    hydrodynamical simulation is the same 2D multi-spot ignition
    delayed detonation model for all cases.  The left column is for
    tracer particles of constant mass, whereas for the right column
    the variable tracer mass approach was used.  The radius, $s_i$, of
    the markers increases with mass fraction $X_i$ according to
    $s_i=\max\{0.1,\,29.9[\log_{10}(X_i)+5]/5+0.1\}$ in arbitrary
    units. Red means overproduction, blue means underproduction, and
    green means good agreement with respect to the case with the most
    tracer particles.}
  \label{fig3}
\end{figure*}

We relax the widely-used constraint of equal mass tracer particles and
implement a mass coordinate dependent distribution (more lighter
particles at low density, see Fig.~\ref{fig1}).  These variable mass
tracers better resolve the morphology of the low density region where
intermediate mass elements form (see Fig.~\ref{fig2}). Furthermore,
the convergence with particle number of the integrated isotopic
nucleosynthetic yields of some intermediate mass isotopes can be
improved (see Fig.~\ref{fig3}).

\section{Conclusion}
The tracer particle method yields appear to begin to converge for
tracer particle numbers greater than $\sim$32 per axis and direction
(see Fig.~\ref{fig3}). The yield convergence for nucleosynthetic
products of incomplete burning, such as e.g. Mg or Al, can be improved
upon by using tracer particles of variable mass. These variable mass
tracers can greatly improve the spatial resolution of the lower
density nucleosynthetic yield distribution, which could lower the
number of particles required to obtain converged light curves and
spectra (see e.g. \cite{kromer2009a}). For further details, please see
the journal article associated with this poster
\cite{seitenzahl2010a}.

\acknowledgments{This work was supported by the Deutsche
  Forschungsgemeinschaft via Emmy Noether Program (RO 3676/1-1) and
  the Excellence Cluster EXC~153.}

\bibliographystyle{./PoS} \bibliography{./astrofritz}


\end{document}